# Matching Scherrer's k essence argument with behavior of scalar fields permitting derivation of a cosmological constant

A. W. Beckwith

Department of Physics and Texas Center for Superconductivity and Advanced Materials at the University of Houston
Houston, Texas 77204-5005 USA

## ABSTRACT

We previously showed that we can use particle-antiparticle pairs as a model of how nucleation of a new universe occurs. We now can construct a model showing evolution from a dark matter dark energy mix to a pure cosmological constant cosmology due to changes in the slope of the resulting scalar field, using much of Scherrer's k-essence model. This same construction permits a use of the speed of sound, in k essence models evolving from zero to one. Having the sound speed eventually reach unity permits matching conventional cosmological constant observations in the aftermath of change of slope of a S-S' pair during the nucleation process of a new universe. This also assumes that Scherrer's derivation of a sound speed being zero is appropriate during initial inflationary cosmology.

Correspondence: A. W. Beckwith:    projectbeckwith2@yahoo.com

# INTRODUCTION

We[1] have investigated the role an initial false vacuum procedure with a driven sine Gordon potential plays in the nucleation of a scalar field in inflationary cosmology. Here, we show how that same scalar field blends naturally into the chaotic inflationary cosmology presented by Guth,[2] which has its origins in the evolution of nucleation of an electron-positron pair in a de Sitter cosmology. The final results of this model, when $\phi \to \varepsilon^+$, appears congruent with the existence of a region that matches the flat slow roll requirement of $\left|\frac{\partial^2 V}{\partial \phi^2}\right| << H^2$; the negative pressure requirement involving both first and second derivatives of the potential w.r.t. scalar fields divided by the potential itself being very small quantities, where $H$ is the expansion rate that is a requirement of realistic inflation models.[4] This is due to having the potential in question $V \propto \phi^2 \xrightarrow[\phi \to \varepsilon^+]{} V_0 \equiv$ constant for declining scalar values.

We have formed, using Scherrer's argument,[3] a template for evaluating initial conditions to shed light on whether this model universe is radiation-dominated in the beginning or is more in sync with having its dynamics determined by assuming a straight cosmological constant. Our surprising answer is that we do not have conditions for formation of a cosmological constant-dominated era when close to a thin wall approximation of a scalar field of a nucleating universe, but that this is primarily due to an extremely sharp change in slope of the would-be potential field . The sharpness of this slope, leading to a near delta function behavior for kinematics at the thin wall approximation for the initial conditions of an expanding universe would lead, at a later time, to conditions appropriate for necessary and sufficient cosmological dynamics largely controlled by a cosmological constant when the scalar field itself ceases to be affected by the thin wall approximation but is a general slowly declining slope.

## HOW DARK MATTER TIES IN, USING PURE KINETIC K ESSENCE AS DARK MATTER TEMPLATE FOR A NEAR THIN WALL APPROXIMATION OF $\phi$

We define k essence as any scalar field with non-cannonical kinetic terms. Following Scherrer,[3] we introduce a momentum expression via

$$p = V(\phi) \cdot F(X) \tag{1}$$

where we define the potential in the manner we have stated for our simulation as well as set[3]

$$X = \frac{1}{2} \cdot \nabla_\mu \phi \ \nabla^\mu \phi \tag{2}$$

and use a way to present $F$ expanded about its minimum and maximum[3]

$$F = F_0 + F_2 \cdot (X - X_0)^2 \tag{3}$$

where we define $X_0$ via $F_X\big|_{X=X_0} = \frac{dF}{dX}\bigg|_{X=X_0} = 0$, as well as use a density function[3]

$$\rho \equiv V(\phi) \cdot [2 \cdot X \cdot F_X - F] \tag{4}$$

where we find that the potential neatly cancels out of the given equation of state so[3]

$$w \equiv \frac{p}{\rho} \equiv \frac{F}{2 \cdot X \cdot F_X - F} \tag{5}$$

as well as a growth of density perturbations terms factor Garriga and Mukhanov[4] wrote as

$$C_x^2 = \frac{(\partial p / \partial X)}{(\partial \rho / \partial X)} \equiv \frac{F_X}{F_X + 2 \cdot X \cdot F_{XX}} \tag{6}$$

where $F_{XX} \equiv d^2 F / dX^2$, and since we are fairly close to an equilibrium value, we pick a value of X close to an extremal value of $X_0$.[3]

$$X = X_0 + \tilde{\varepsilon}_0 \tag{7}$$

where, when we make an averaging approximation of the value of the potential due to Fig. 1b as very approximately a constant, we may write the equation for the k essence field as taking the form (where we assume $V_\phi \equiv dV(\phi)/d\phi$ )[3]

$$(F_X + 2 \cdot X \cdot F_{XX}) \cdot \ddot{\phi} + 3 \cdot H \cdot F_X \cdot \dot{\phi} + (2 \cdot X \cdot F_X - F) \cdot \frac{V_\phi}{V} \equiv 0 \tag{8}$$

as approximately

$$(F_X + 2 \cdot X \cdot F_{XX}) \cdot \ddot{\phi} + 3 \cdot H \cdot F_X \cdot \dot{\phi} \cong 0 \tag{9}$$

which may be re written as[3]

$$(F_X + 2 \cdot X \cdot F_{XX}) \cdot \ddot{X} + 3 \cdot H \cdot F_X \cdot \dot{X} \cong 0 \tag{10}$$

In this situation, this means that we have a very small value for the growth of density pertubations[3]

$$C_s^2 \cong \frac{1}{1 + 2 \cdot (X_0 + \tilde{\varepsilon}_0) \cdot (1/\tilde{\varepsilon}_0)} \equiv \frac{1}{1 + 2 \cdot \left(1 + \frac{X_0}{\cdot \tilde{\varepsilon}_0}\right)} \tag{11}$$

when we can approximate the *kinetic energy* from

$$(\partial_\mu \phi) \cdot (\partial^\mu \phi) \equiv \left(\frac{1}{c} \cdot \frac{\partial \phi}{\partial \cdot t}\right)^2 - (\nabla \phi)^2 \cong -(\nabla \phi)^2 \rightarrow -\left(\frac{d}{dx} \phi\right)^2 \tag{12a}$$

and if we assume that we are working with a comparatively small contribution w.r.t. time variation but a very large, in many cases, contribution w.r.t. spatial variation of phase

$$|X_0| \approx \frac{1}{2} \cdot \left(\frac{\partial \phi}{\partial x}\right)^2 >> \tilde{\varepsilon}_0 \tag{12b}$$

$$0 \leq C_S^2 \approx \varepsilon^+ << 1 \tag{13}$$

and[3,5]

$$w \equiv \frac{p}{\rho} \cong \frac{-1}{1 - 4 \cdot (X_0 + \tilde{\varepsilon}_0) \cdot \left(\frac{F_2}{F_0 + F_2 \cdot (\tilde{\varepsilon}_0)^2} \cdot \tilde{\varepsilon}_0\right)} \approx 0 \tag{14}$$

We get these values for the phase being nearly a box of height approximately scaled to be about $2 \cdot \pi$ and of width **L**. Which we obtained by setting [1]

$$\varphi \approx \pi \cdot [\tanh b \cdot (x + L/2) - \tanh b \cdot (x - L/2)] \tag{15}$$

This means that the initial conditions we are hypothesizing are in line with the equation of state conditions appropriate for a cosmological constant but near zero effective sound speed. As it is, we are approximating

[Insert Fig. 1a and 1b about here]

$$|X_0| \approx \frac{1}{2} \cdot \left(\frac{\partial \phi}{\partial x}\right)^2 \cong \frac{1}{2} [\delta_n^2(x + L/2) + \delta_n^2(x - L/2)] \tag{16}$$

with

$$\delta_n(x \pm L/2) \xrightarrow[n \to \infty]{} \delta(x \pm L/2) \tag{17}$$

as the slope of the S-S' pair approaches a box wall approximation in line with thin wall nucleation of S-S' pairs being in tandem with $b -$ *larger*. Specifically, in our simulation, we had $b - 10$ above, rather than go to a pure box style representation of

S-S' pairs; this could lead to an unphysical situation with respect to delta functions giving infinite values of infinity, which would force both $C_s^2$ and $w \equiv \dfrac{p}{\rho}$ to be zero for

$$|X \approx X_0| \cong \frac{1}{2} \cdot \left(\frac{\partial \phi}{\partial x}\right)^2 \to \infty$$

if the ensemble of S-S' pairs were represented by a pure thin wall approximation,[1] i.e., a box. If we adhere to a finite but steep slope convention to modeling both $C_s^2$ and $w \equiv \dfrac{p}{\rho}$, we get the following: When $b \geq 10$ we obtain the conventional results of

$$w \cong \frac{-1}{1 - 4 \cdot \dfrac{X_0 \cdot \tilde{\varepsilon}_0}{F_2}} \to -1 \qquad (18)$$

and recover Scherrer's solution for the speed of sound[3]

$$C_S^2 \approx \frac{1}{1 + 4 \cdot X_0 \left(1 + \dfrac{X_0}{2 \cdot \tilde{\varepsilon}_0}\right)} \to 0 \qquad (19)$$

(if an example $F_2 \to 10^3$, $\tilde{\varepsilon}_0 \to 10^{-2}$, $X_0 \to 10^3$). Similarly, we would have if $b - 3$ in Eq. (12a) above

$$w \cong \frac{-1}{1 - 4 \cdot \dfrac{X_0 \cdot \tilde{\varepsilon}_0}{F_2}} \to -1 \qquad (20)$$

and

$$C_S^2 \approx \frac{1}{1 + 4 \cdot X_0 \left(1 + \dfrac{X_0}{2 \cdot \tilde{\varepsilon}_0}\right)} \to 1 \qquad (21)$$

if $F_2 \to 10^3$, $\tilde{\varepsilon}_0 \to 10^{-2}$. Furthermore $|X_0| \to$ *a small value*, which for $b - 3$ in Eq. (12a) would lead to $C_S^2 \approx 1$, i.e., when the wall boundary of a S-S' pair is no longer approximated by the thin wall approximation. This eliminates having to represent the initial state as behaving like pure radiation state (as Cardone[5] et al postulated), i.e., we then recover the cosmological constant. When $|X_0| \approx \frac{1}{2} \cdot \left(\frac{\partial \phi}{\partial x}\right)^2 >> \tilde{\varepsilon}_0$ no longer holds, we can have a hierarchy of evolution of the universe as being first radiation dominated, then dark matter, and finally dark energy.

$$\text{If } |X \approx X_0| \cong \frac{1}{2} \cdot \left(\frac{\partial \phi}{\partial x}\right)^2 \to \infty,$$ neither limit leads to a physical simulation that makes sense; so, in this problem, we then refer to the contributing slope as always being large but not infinite. We furthermore have, even with $w = -1$

$$C_s^2 \equiv 1 \xrightarrow[b1 \to 3]{} 1 \quad 3$$

[Insert Figures 2a and 2b about here]

indicating that the evolution of the magnitude of the phase $\phi \to \varepsilon^+$ corresponds with a reduction of our cosmology from a dark energy dark matter mix to the more standard cosmological constant models used in astrophysics.

## CONCLUSION

We have a situation for which we can postulate an early universe which is *not* necessarily radiation dominated as postulated by Carbone et al.[5] We should keep in mind that Scherrer was looking for very small $\varepsilon_1$ and a constant $a_1 > a$, with $a$ written as an expansion scale factor.

$$X = X_0 \cdot \left(1 + \varepsilon_1 \cdot \left(\frac{a}{a_1}\right)^{-3}\right) \tag{22}$$

so he could then get a *general* solution of

$$1 \gg C_x^2 \equiv (X - X_0)/(3 \cdot X - X_0) \equiv \frac{1}{2} \cdot \varepsilon_1 \cdot (a/a_1)^{-3} \approx \varepsilon^+ \geq 0 \tag{23}$$

while at the same time keeping $w = -1$.

However, Scherrer[3] does not take into consideration whether the dark matter dark energy regime is primarily dominant at a given time in cosmological evolution — and throws out the positive cosmological constant all together. Second, Scherrer's model[3] does not take into consideration whether cosmic inflation was dominated by the dark matter dark energy mix in the beginning. I argue that having such a mixture of dark matter dark energy in cosmic expansion would be the driving force in order to establish the cosmic expansion parameters as we know them.

In addition, our kinetic model can be compared with the very interesting Chimentos[6] purely kinetic k–essence model, with density fluctuation behavior at the initial start of a nucleation process. The model indicates our density function reach $\rho =$ constant after passing through the tunneling barrier as mentioned in the first papers nucleation of a S-S' pair ensemble. Topological arguments blends the k essence results indicating Scherrer's dark energy dark matter mixture[3] during the inflationary cosmological period to the decay of the thin wall approximation of the scalar field to conditions permitting the dominant contribution of the cosmological constant to present changes in the Hubble parameter.

## FIGURE CAPTIONS

**Fig 1a, 1b:** Evolution of the phase from a thin wall approximation to a more nuanced thicker wall approximation with increasing L between S-S' instanton componets. The height drops and the width L increases corresponds to a de evolution of the thin wall approximation. This is in tandem with a collapse of an initial nucleating potential system to Guth's standard chaotic scalar $\phi^2$ potential system. As the curve flattens, and the thin wall approximation dissipates, the physical system approaches standard cosmological constant behavior.

**Fig. 2a, 1b**: As the *walls* of the S-S' pair approach the thin wall approximation, one finds that for a normalized distance $L = 9 - \boldsymbol{L} = 6 - L = 3$ that one has an approach toward delta function behavior at the boundaries of the new, nucleating phase. As $L$ increases, the delta function behavior subsides dramatically. Here, the $L = 9 \Leftrightarrow$ conditions approach a cosmological constant; $\boldsymbol{L} = 6 \Leftrightarrow$ conditions reflect Sherrer's dark energy dark matter mix; $L = 3 \Leftrightarrow$ approach unphysical delta function contributions due to a pure thin wall model.

**Figure 1a, 1b
Beckwith**

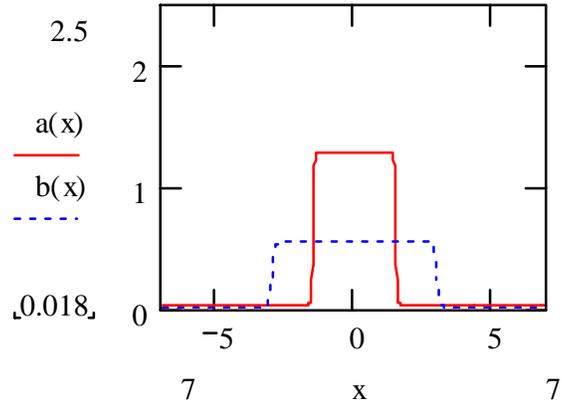

**1a**

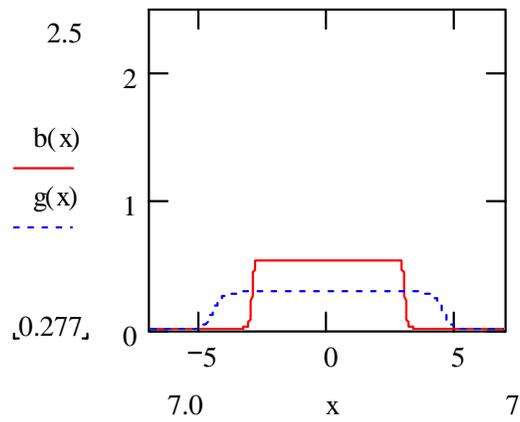

**1b**

**Figure 2a, 2b
Beckwith**

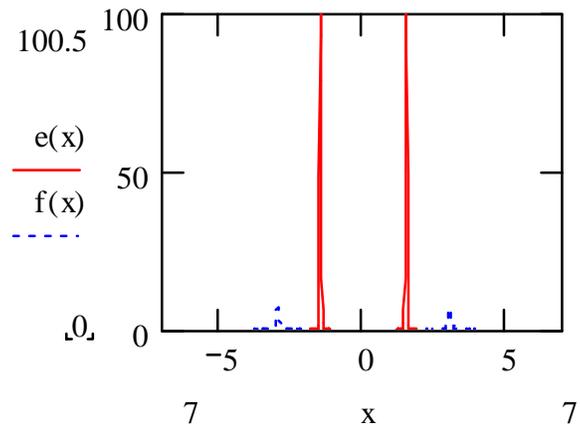

**2a**

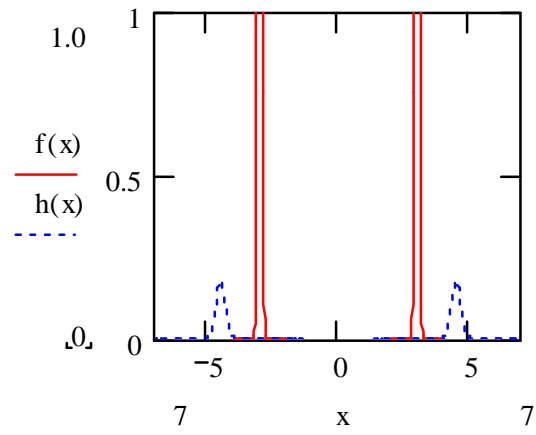

**2b**